\documentclass[a4paper,10pt]{article}
\usepackage[utf8]{inputenc}
\usepackage{amsmath,amssymb,amsfonts}
\usepackage{graphicx}
\usepackage{textcomp}
\usepackage{xcolor}
\usepackage{float}
\usepackage{longtable}
\usepackage{array}
\usepackage{ragged2e}
\usepackage{tabularx}
\usepackage[hyphens]{url}
\usepackage{hyperref}

\title{Green by Design? Investigating the Energy and Carbon Footprint of Chia Network}

\author{Soraya Djerrab, ESTIN, Algeria, s\_djerrab@estin.dz \\
Cl\'ementine Gritti, INSA Lyon - Inria, France, clementine.gritti@insa-lyon.fr \\
Rahima Benzenati, ESTIN, Algeria, benzenatirahima@estin.dz}

\begin{document}

\maketitle

\begin{abstract}
This paper presents a detailed analysis of the environmental impact of Chia Network (Chia for short), a green-claimed blockchain, which uses a Proof of Space and Time (PoST) consensus mechanism. While Chia claims to be a sustainable alternative to Proof-of-Work-based blockchains, our results show that its resource-intensive initialization phase and ongoing operations lead to carbon emissions 18× higher than claimed (0.88 MtCO\textsubscript{2}/year), exceeding mainstream "green" blockchains by orders of magnitude. We combine experimental measurements from a controlled testbed (Grid’5000) with theoretical modeling of operational and embodied emissions to assess Chia's true sustainability profile.
\end{abstract}

\section{Introduction}
The rapid growth of blockchain technology has raised significant concerns about its environmental impact, particularly due to energy-intensive consensus mechanisms like Proof of Work (PoW). While alternatives such as Proof of Stake (PoS) have emerged as more sustainable options, Chia proposes a different approach with its Proof of Space and Time (PoST) consensus, claiming to be an eco-friendly alternative by leveraging storage rather than computational power. However, the true environmental footprint of Chia remains understudied, with its reliance on storage-intensive plotting and farming introducing new sustainability trade-offs.

Global carbon neutrality initiatives like the European Green Deal \cite{eugreendeal} demand sustainable technologies. Yet Bitcoin alone emits 92.2 MtCO\textsubscript{2} \cite{ccaf2025bitcoin,KOHLI202379} - comparable to entire nations. While "green" alternatives like PoS reduce operational energy, Chia's storage-dependent design risks shifting environmental burdens to hardware production and accelerated wear. This necessitates rigorous assessment of its true lifecycle impacts.

This paper investigates the energy and carbon footprint of Chia, critically assessing its claims of sustainability. We combine empirical measurements from controlled experiments with theoretical modeling to evaluate both operational energy consumption and embodied carbon emissions from hardware manufacturing and wear. Our analysis reveals that Chia’s footprint is significantly higher than claimed, exceeding that of leading PoS blockchains by orders of magnitude.

Our key contributions include:
\begin{itemize}
    \item Empirical energy measurements of plotting and farming phases under different configurations, using a controlled testbed (Grid’5000).
    \item Two complementary modeling approaches—homogeneous scaling and cohort-based analysis—to estimate Chia’s annual carbon emissions.
    \item A sensitivity analysis exploring how hardware distribution and plot compression adoption impact the results.
    \item A critical comparison of Chia’s footprint against other "green" blockchains, highlighting discrepancies in sustainability claims.
\end{itemize}
Our findings highlight the importance of rigorous, transparent, and empirically grounded assessments when evaluating the sustainability claims of emerging blockchain technologies.
\section{Related Work}
\subsection{Green Blockchains and Consensus Mechanisms}
The environmental impact of blockchain technology gained major attention since the emergence of Bitcoin in 2009 \cite{nakamoto2009}, mainly due to its Proof-of-Work (PoW) consensus. PoW mechanism requires nodes (miners) to solve computationally intensive puzzles to validate transactions. This process is estimated to consume 92.2 MtCO\textsubscript{2} per year  -- comparable to Morocco's annual electricity use \cite{ccaf2025bitcoin,KOHLI202379}. Ethereum, which traditionally used PoW, transitioned to Proof-of-Stake (PoS) in 2022 for environmental reasons. PoS is a consensus mechanism where validators are chosen based on staked cryptocurrency rather than computational power. The transition reduced Ethereum's energy demand by 99.99\% \cite{greenblockchains}, highlighting the role of consensus design in sustainability. Alternative "green" blockchains like Algorand (Pure PoS), Tezos (Liquid PoS), Celo (PoS with a mobile approach), Cardano (Ouroboros, a member of the PoS family), Polkadot (Nominated Proof of Stake), Avalanche (Avalanche consensus, a PoS variant) , Solana (PoS), and many more  further demonstrate lower operational energy (less than 0.001 TWh/year \cite{ccri}) by eliminating computational competition typical of PoW \footnote{\url{https://ethereum.org}, \url{https://algorand.co/}, \url{https://tezos.com}, \url{https://celo.org/}, \url{https://cardano.org}, \url{https://polkadot.network}, \url{https://www.avax.network/}, \url{https://solana.com/}}.

Chia introduced Proof-of-Space-and-Time (PoST) \cite{chiahome}, replacing computation with storage-based validation. While claiming energy efficiency, Chia shifts carbon emissions towards new sources, which will be further discussed in this paper.

\subsection{Previous Work}

Recent research has explored the environmental footprint of blockchain technologies, both in terms of identifying low-impact alternatives and improving energy estimation methods. Alzoubi and Mishra \cite{greenblockchains} conducted a hybrid literature review to identify 23 blockchain networks with significantly lower energy consumption and carbon emissions than Bitcoin, including Algorand, Cardano, Tezos, and Bitgreen. However, their study lacked empirical measurements and called for more data-driven evaluations. Algorand’s own methodology \footnote{\url{https://algorandtechnologies.com/news/sustainable-blockchain-calculating-the-carbon-footprint}} estimates the energy per finalized transaction using efficient hardware (e.g., Raspberry Pi 4), but simplifies many factors and only considers validator nodes. In contrast, several studies focus on Bitcoin and criticize the reliability of existing energy models. For example, the Cambridge Bitcoin Electricity Consumption Index (CBECI) \cite{ccaf2025bitcoin} uses average hash rates to estimate power usage, while Digiconomist’s index \cite{digiconomist2024} relies on a revenue-based approach, assuming electricity costs represent a stable portion of miner income. Krause and Tolaymat’s influential bottom-up model \cite{article}, which was also implemented in \cite{gallersdorfer2020energy}, estimates energy use by dividing network hash rates by device efficiency per algorithm, but also acknowledges major uncertainties due to the opaque nature of mining infrastructure. Reviews by Alshahrani et al. \cite{en16031510}, Sai and Vranken \cite{SAI2024100169}, and Lei et al. \cite{LEI2021112422} highlight widespread issues across the field, including unjustified assumptions, lack of transparent data and code, outdated parameters, and weak empirical grounding. The systematic review by Sai et al. \cite{SAI2024100169} reveals foundational gaps: 74\% of the analyzed studies do not build upon existing theories in blockchain or any other related domain, 34\% of studies lack explicit research design, 43\% withhold data, 67\% do not share code, and 79\% fail to discuss external data reliability. These reviews advocate for rigorous methodological standards, including open data, clear system boundaries, and empirical validation.

\section{Background}

\subsection{Overview of Chia and PoST}
\begin{figure}[H]
         \centering
         \includegraphics[width=1\linewidth]{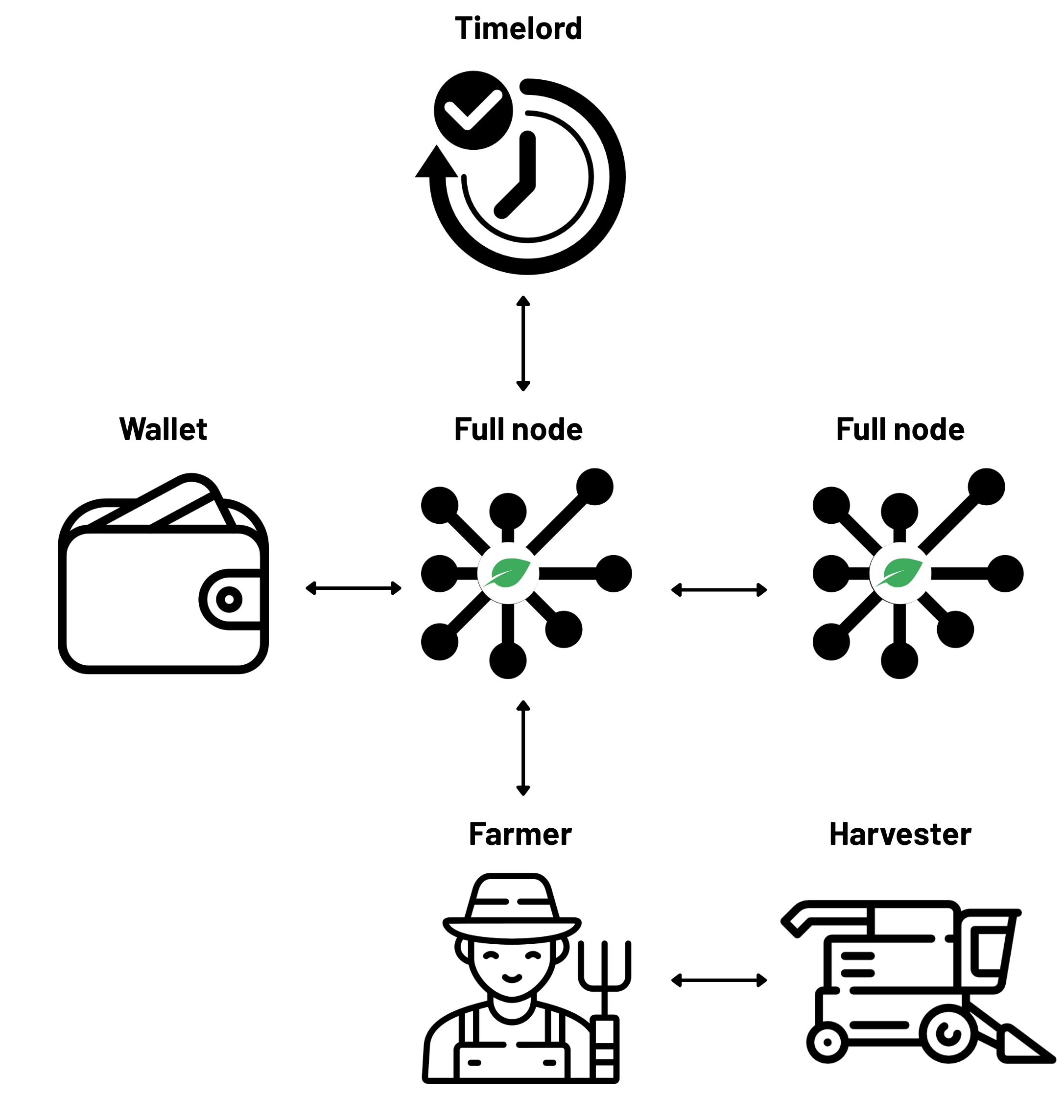}
         \caption{Chia's architecture overview}
         \label{fig:chia structure}
    \end{figure}

Chia is a decentralized smart‑transaction platform founded by Bram Cohen and Ryan Singer in 2017. Rather than relying on energy‑intensive Proof of Work (PoW), Chia secures its network via Proof of Space and Time (PoST) -- a two-part consensus that harnesses unused disk capacity and cryptographic delays to validate blocks. \cite{chiadocumentation} \cite{chiahome}. Figure \ref{fig:chia structure} shows the basic architecture for Chia.
\begin{itemize}
    \item \textbf{Proof of Space} : Farmers pre-compute and store large cryptographic files called plots (108.8 GiB for k=32) on disk. When the network issues a challenge, each farmer’s harvester scans these plots for a valid proof \cite{Chiagreenpaper}.
    \item \textbf{Proof of Time} : To prevent grinding attacks (rapid, repeated attempts to find proofs), a specialized node called the timelord computes a Verifiable Delay Function (VDF) \cite{boneh2018verifiable} -- a sequential cryptographic operation that enforces a minimum, un‑parallelizable time delay before a new block can be created \cite{Chiagreenpaper}
\end{itemize}
These components are orchestrated by four core processes (which often run on the same machine) \footnote{\url{https://docs.chia.net/architecture-overview/}}:
\begin{itemize}
    \item \textbf{Full Node:} Maintains and validates the blockchain.
    \item \textbf{Harvester:} Scans local plots in response to challenges.
    \item \textbf{Farmer:} Receives proofs from harvesters, verifies them, and submits winning proofs for block creation.
    \item \textbf{Wallet:} Manages private keys and collects XCH rewards.
\end{itemize}
\subsubsection{Plotting}
Plotting is the resource-intensive phase where farmers generate on-disk plots for farming. The standard Chia plotter takes 6--12 hours per file on consumer hardware, writing 239 GiB of temporary data to SSDs and causing significant wear. Optimized plotters mitigate this: MadMax uses multi-threaded CPU to reduce times to under 30 minutes on high-end machines; Bladebit provides \texttt{In-RAM} (RAM-only, needing 416 GiB minimum) and \texttt{CUDA} (GPU-accelerated) modes; and Gigahorse, an unofficial suite, supports high-compression formats for parallel plotting and farming\footnote{\url{https://github.com/madMAx43v3r/chia-plotter}, \url{https://github.com/Chia-Network/bladebit}, \url{https://github.com/madMAx43v3r/chia-gigahorse}}. Compressed plots ( which are smaller incomplete plots, e.g., 78 GiB for C7 level) reduce disk usage but increase farming CPU/memory demands due to the need for reconstructing the plots when checking for proofs.
\subsubsection{Farming}
Farming is the phase that follows plotting. It involves the process by which a farmer node continuously responds to cryptographic challenges issued by Chia, approximately every 18.75 seconds \cite{Chiagreenpaper}. Each challenge is used to check existing plot files for potential valid proofs, in which case the farmer creates the block and wins a reward.

\subsection{Energy Evaluation Methodologies}
Blockchain energy estimation employs two primary approaches:
\begin{itemize}
    \item \textbf{Bottom-up models} estimate the electricity consumption of blockchain networks by summing the energy use of all individual computing devices. Accurate estimation requires knowledge of the type and quantity of hardware used in the network, along with device-specific metrics such as energy consumption and uptime. The total electricity consumption \(T\) of the network can be calculated as \(T = \sum \varepsilon(i) \cdot PUE(i)\), where \(\varepsilon(i)\) is the energy consumption of device \(i\), and \(PUE(i)\) accounts for additional operational overhead such as cooling and networking \cite{SAI2024100169}.
    \\
    Studies of PoS systems (e.g., Cardano, Solana) combine validator counts, hardware profiles, and Power Usage Effectiveness (PUE) adjustments \cite{ibañez2023energyconsumptionproofofstakesystems}.
    Sai et al.~\cite{SAI2024100169} emphasize that such approaches, when executed rigorously, offer more accurate insights into the energy demands of blockchain systems.
    \item \textbf{Top-down models} : In decentralized blockchain systems such as Bitcoin, the lack of transparency regarding the geographic distribution and hardware used by participants (miners) makes bottom-up energy estimation challenging. As a result, many studies adopt a top-down modeling approach, which estimates energy consumption using macro-level variables like hash rate, economic incentives, and electricity pricing.
    \\
    Top-down models do not require exact hardware data. Instead, they rely on the assumption that the total network hash rate \(H\) is generated by a hypothetical pool of devices, each with known energy and performance characteristics. This hash rate can be inferred from the mining difficulty and average block time (e.g., 10 minutes for Bitcoin) \cite{SAI2024100169}.
    The total electricity consumption \(T\) of the network can be calculated as \cite{LEI2021112422} :
    \(T = H \cdot e \cdot PUE\), where \(H\) is hash rate (hash/second), \(e\) is energy efficiency of the hardware (J/hash), and \(PUE\) is power usage effectiveness.
\end{itemize}
\section{Experimental setup and results}
\subsection{Testbed: Grid'5000}
All experiments were conducted on the Grid’5000 experimental testbed, a large-scale and flexible platform for research in distributed systems, cloud computing, and HPC \footnote{\url{https://www.grid5000.fr/w/Grid5000:Home/}}. We used the \textbf{Gemini} \footnote{\url{https://www.grid5000.fr/w/Lyon:Hardware\#gemini}} node in the Lyon site, which is equipped with:
\begin{itemize}
    \item \textbf{Model:} Nvidia DGX-1
    \item \textbf{CPU:} Intel Xeon E5-2698 v4 (Broadwell), x86\_64, 2.20GHz, 2 CPUs/node, 20 cores/CPU
    \item \textbf{Memory:} 512 GiB
    \item \textbf{Storage:} disk0, 480 GB SSD SATA Samsung SAMSUNG MZ7KM480
    \item \textbf{GPU:} 8 x Nvidia Tesla V100-SXM2-32GB (32 GiB), Compute capability: 7.0
\end{itemize}

\subsection{Monitoring Tools}

To measure energy consumption and system-level resource usage in a reproducible and rigorous manner, we relied on a combination of software and hardware monitoring tools, integrated and automated using a unified Bash script. This script launched the plotting or farming operation in the background and initiated real-time data collection through multiple tools, logging energy and I/O metrics for post-processing. After each run, raw data was parsed using shell and Python scripts to extract total energy usage (Wh) and disk writes (TiB).

\begin{itemize}
    \item \textbf{/proc/diskstats}: is a Linux kernel interface that provides accurate, low-level statistics on disk I/O activity, such as the number of sectors written and time spent writing. It operates directly at the block layer, making it ideal for measuring real disk usage and estimating SSD wear during intensive operations like plotting \cite{iotop}.
    \item \textbf{kwollect wattmeter:} Kwollect is a hardware-level energy monitoring system used on Grid’5000. It collects high-frequency power data from wattmeters attached to compute nodes and provides accurate energy usage measurements via API and dashboards. Its precision and granularity make it highly reliable for energy analysis in experiments \cite{kwollect}.
\end{itemize}

\subsubsection{Measurement Selection Justification}
Although multiple monitoring tools were used during experimentation, only the data from \texttt{Kwollect} (for energy) and \texttt{/proc/diskstats} (for disk I/O) were used in our model. Kwollect offers hardware-level accuracy, making it the most reliable for energy consumption analysis. For I/O, Diskstats provides kernel-level precision on sector writes, ideal for estimating SSD writes and thus, SSD wear.

\texttt{Pidstat}, a tool for tracking per-process CPU and I/O usage over time, gave results close to diskstats and was useful when monitoring external drives. \texttt{Iotop}, which shows real-time disk I/O by processes, consistently overestimated I/O in our tests. \texttt{Scaphandre}, an open-source agent that uses Intel RAPL to estimate per-process energy, showed inconsistent results with over- or under-estimations depending on the scenario. \texttt{CodeCarbon}, a Python library that estimates CO\textsubscript{2} emissions from code execution, produced results close to \texttt{Kwollect} but could not be implemented across all experiments, as it requires modification of the source code to track energy use \footnote{\url{https://man7.org/linux/man-pages/man1/pidstat.1.html}, \url{https://man7.org/linux/man-pages/man8/iotop.8.html}, \url{https://github.com/hubblo-org/scaphandre}, \url{https://codecarbon.io/}}.

\subsection{Experimentation Scenarios}
We carried out 10 experiments to evaluate the energy consumption and carbon footprint of both the plotting and farming phases for Chia version 2.5.2 under different configurations. Each experiment was run using the same Bash automation script, ensuring consistent monitoring and controlled environments.
\begin{itemize}
    \item Standard plotting (Chia GUI default) : Represents the baseline plotting method using the original Chia CLI with a k32 plot.
    \item Bladebit plotting (RAM mode) : Uses BladeBit software, and needs a minimum of 512 GiB RAM. It minimizes I/O operations.
    \item BladeBit plotting (GPU mode) : Leverages GPU acceleration (CUDAPlot) for faster plotting performance with BladeBit cudaplot mode.
    \item MadMax plotting : A CPU-optimized faster plotter that improves speed through parallelism and multi-threads.
    \item Farming with standard plots (k32) : Evaluates energy usage during the passive farming phase using traditional, uncompressed plots.
    \item Farming with compressed plots (C5) : Tests the impact of plot compression on energy use during farming.
\end{itemize}

\subsection{Results}

Table~\ref{tab:results} summarizes the key performance and resource usage metrics collected during our experimental scenarios. Each value corresponds to the average obtained from multiple runs per setup (between 6 to 10 runs). We observe that BladeBit in GPU mode yields the fastest plotting time (7.35 minutes) with the lowest energy usage (86 Wh), followed closely by RAM mode (165.6 Wh). In contrast, Chia’s standard plotter consumes over 4.9 kWh per plot, which is nearly 30× more energy than BladeBit GPU and over 5× more than MadMax. Disk I/O also follows the same trend: standard plotting causes around 1.64 TiB of SSD writes per plot, while BladeBit methods reduce this to less than 0.1 TiB, indicating significantly lower SSD wear. For farming, we tested both with compressed and uncompressed plots, but had no noticeable difference, likely due to our small farm -- only 20 plots.

\begin{table*}[htbp]
\centering
\caption{Summary of experimental results}
\label{tab:results}
\begin{tabular}{|l|c|c|c|c|c|c|c|c|}
\hline
\textbf{Experiment} & \textbf{Duration (min)} & \textbf{kwollect (Wh)} & \textbf{Diskstats (TiB)} \\
\hline
Standard Plotting (Chia) & 379.02 & 4995.0485 & 1.64 \\
MadMax Plotting & 64.48 & 927.634 & 1.357\\
Bladebit Plotting (RAM) & 10.955 & 165.637 & 0.083376\\
Bladebit Plotting (GPU) & 7.346 & 85.968 & 0.0813125 \\
Farming  & 60 & 771.836 & 0.000097\\
\hline
\end{tabular}
\end{table*}

\section{Analysis}

We propose two complementary modeling approaches to estimate Chia's yearly energy use.  Method 1 \ref{subsection:first method} uses purely empirical measurements from our server experiments (plotting and farming) and scales them directly to the entire network netspace and node count, applying a single PUE factor for overhead.  Method 2 \ref{subsection:method2} refines this by segmenting the network into three hardware cohorts (servers, desktops, and laptops), each with its own parameters (energy draw, PUE, embodied carbon) and usage shares. This cohort approach captures heterogeneity in both resource efficiency and carbon intensity, allowing us to compare a uniform model (Method 1) against a more granular, device-specific estimate (Method 2).

\subsection{Notation and Common Parameters}

Table \ref{tab:variables} shows the variables used in our models with their definitions, units, and assumed values:

\begin{table*}[htbp]
\centering
\caption{Notations and values}
\label{tab:variables}
\begin{tabular}{|p{0.2\textwidth}|p{0.6\textwidth}|p{0.1\textwidth}|p{0.06\textwidth}|}
\hline
\textbf{Symbol} & \textbf{Definition} & \textbf{Value Used} & \textbf{Source} \\
\hline
$S_{\mathrm{net}}$ (EiB) & Total network netspace & 33.8465 & ** \cite{chianetspace}\\
$S_{\mathrm{netg}}$ (EiB) & Netspace growth & 12.6593 & ** \cite{chianetspace} \\
$N_{\mathrm{node}}$ & Number of nodes in the network & 250000 & ** \cite{chianeilcohen} \\
$S_{\mathrm{plot}}$ (GiB)& Size of a \(k=32\) plot & 101.4 & ** \cite{chiadocumentation}\\
$S_{\mathrm{plot,C5}}$ (GiB) & Size of a \(k=32\) plot compressed C5. & 81.3 & ** \cite{chiadocumentation} \\
$E_{\mathrm{plot}}$ (kWh)& Empirical energy to create one standard (uncompressed) plot. & 4.995 & * \\
$E_{\mathrm{plot,C5,RAM}}$ (Wh) & Empirical energy to create one compressed plot C5 with BladeBit Ram mode. & 165.637 & *\\
$E_{\mathrm{plot,C5,GPU}}$ (Wh) & Empirical energy to create one compressed plot C5 with BladeBit Cuda. & 85.968 & *\\
$E_{\mathrm{plot,MM}}$ (Wh) & Empirical energy to create one standard (uncompressed) plot with MadMax plotter & 927.634 & *\\
$E_{\mathrm{farm}}$ (kWh) & Empirical annual farming energy for one farmer with uncompressed plots & 6761.283 & *  \\
$PUE$ (unitless) & Power Usage Effectiveness of the infrastructure for servers. & 1.58 & ** \cite{pue} \\
$I_{\mathrm{elec}}$ (kg CO\textsubscript{2}/kWh) & Grid carbon intensity. & 0.384 & ** \cite{carbonintensity} \\
$T_{\mathrm{writes}}$(TiB) & Number of writes occuring on SSD during creation of 1 plot with the standard Chia plotter. & 1.64 & * \\
$T_{\mathrm{writes,MM}}$ (TiB) & Number of writes occuring on SSD during creation of 1 plot with the MadMax Chia plotter & 1.357 & * \\
$T_{\mathrm{writes,BB}}$ (TiB) & Number of writes occuring on SSD during creation of a C5 compressed plot with BladeBit & 0.084 & * \\
$\gamma_{\mathrm{SSD}}$ (kgCO\textsubscript{2}e/TiB) & Embodied carbon of manufacturing 1 TiB of SSD. & 160 & ** \cite{Tannu_2023} \\
$\gamma_{\mathrm{HDD}}$ (Kg CO\textsubscript{2}e/TiB) & Embodied carbon of manufacturing 1 TiB of HDD. & 20 & ** \cite{Tannu_2023} \\
$\gamma_{\mathrm{GPU}}$ (kgCO\textsubscript{2}e/GPU)& Embodied carbon emissions for one GPU & 200 & ** \cite{pr13020595} \\
$\gamma_{\mathrm{enter}}$ (kg CO\textsubscript{2}) & Embodied carbon values (CAPEX) for one server & 1000 & ** \cite{pr13020595}\\
$TBW_{\mathrm{SSD}}$ (TiB)& Total bytes written rating of the SSD for a server. & 2390.15207 & ** \cite{tbwenter} \\
$L_{\mathrm{lifetime}}$ (yr) & Expected operational lifetime of hardware components & 4 & ** \cite{lifetiemeworkstation} \\
$f_{\mathrm{BB,enter}}$ & Fraction of servers' netspace in C5 (compressed) plots for servers, computed with BladeBit & 0.6  & ** \cite{xchcompressednetspace}\\
$f_{\mathrm{MM,enter}}$ & Fraction of non-C5 plots created with MadMax for servers & 0.3 & **\\
$f_{\mathrm{std,enter}}$ & Fraction of non-C5 plots created with the standard plotter for servers & 0.1 & **\\
$f_{\mathrm{allocation}}$ & Allocation percentage of Chia's use on a computer. & 0.67 & **  \\
\hline
\multicolumn{4}{l}{\small *: Obtained from our experiments}\\
\multicolumn{4}{l}{\small **: Value from literature or assumed.}
\end{tabular}
\end{table*}
\subsection{Justifications, assumptions and disclaimers:}
For total netspace ($S_{\mathrm{net}}$), we took the peak value at 2024/02/01 from \cite{chianetspace}, meaning our calculations only account for netspace up to this date. Netspace growth ($S_{\mathrm{netg}}$) was derived as the delta between 2023/02/01 and 2024/02/01 from the same source. We use the total netspace to account for embodied storage emissions, and the netspace growth to account for the energy needed to plot for a year. The average carbon intensity ($I_{\mathrm{elec}}$) uses USA's 2024 value from \cite{carbonintensity}, assuming most farmers are US-based. Empirical values from our experiments include $E_{\mathrm{plot}}$, $E_{\mathrm{plot,C5,RAM}}$, $E_{\mathrm{plot,C5,GPU}}$,$E_{\mathrm{plot,MM}}$, $E_{\mathrm{farm}}$, $T_{\mathrm{writes}}$, $T_{\mathrm{writes,MM}}$, and $T_{\mathrm{writes,BB}}$. The SSD's TBW ($TBW_{\mathrm{SSD}}$) is the actual TBW of the SSD used during our experiments \cite{tbwenter}, while $L_{\mathrm{lifetime}}$ (a literature-standard value used in \cite{lifetiemeworkstation}) assumes uniform component lifetimes. For plotting methods: compressed plots (BladeBit) are resource-intensive but efficient, so we assume 60\% of server-equipped farmers use it ($f_{\mathrm{BB,enter}}$ from \cite{xchcompressednetspace}); MadMax ($f_{\mathrm{MM,enter}}$) is assigned to 30\% of such farmers as a mid-tier option; and standard plotting ($f_{\mathrm{std,enter}}$) is deemed unlikely for high-resource farmers. Embodied emissions allocation ($f_{\mathrm{allocation}}$) uses time-based partitioning (66.67\% for Chia) assuming 8-hour daily computer use for other tasks. Storage embodied carbon values are $\gamma_{\mathrm{SSD}}$ and $\gamma_{\mathrm{HDD}}$ (from \cite{Tannu_2023}), while server embodied carbon are $\gamma_{\mathrm{enter}}$ ( value greatly in literature \cite{ji2024scarifcarbonmodelingcloud} \footnote{\url{https://adwdevelopments.com/sustainability/technical-information-paper-embodied-carbon-in-enterprise-data-centre-it-equipment/}}, we choose the value from \cite{pr13020595} as a medium one). We assume all HDDs/SSDs have 1 TiB capacity, plots are created on SSDs and stored on HDDs, and among BladeBit users, half employ GPU plotting while half use RAM plotting.

\subsection{First method} \label{subsection:first method}
This method scales empirical per-plot/farming energy measurements to the total network netspace ($S_{\mathrm{net}}$) and annual netspace growth ($S_{\mathrm{netg}}$). Plotting energy is split between MadMax, BladeBit, and the standard plotter. Embodied carbon for SSDs is based on writes during plotting, while HDDs are scaled by stored netspace. Server emissions subtract pre-included storage to avoid double counting.

\subsubsection{Specific notation}
Netspace share for compressed (C5) plots $S_{\mathrm{C5}}$ = 796,454,2.894 TiB , plots created with Chia's standard plotter $S_{\mathrm{std}}$ = 132,742,3.815 TiB, and plots created with MadMax $S_{\mathrm{MM}}$ = 398,227,1.447 TiB.
\[
  S_{\mathrm{C5}} = f_{\mathrm{BB,enter}}\,S_{\mathrm{netg}},
\]
\[
  S_{\mathrm{std}} = f_{\mathrm{std,enter}}\,S_{\mathrm{netg}},
  \quad
  S_{\mathrm{MM}} = f_{\mathrm{MM,enter}}\,S_{\mathrm{netg}}.
\]
Number of C5 compressed plots $N_{\mathrm{plot,C5}}$ = 100,316,014.136 plots, number of plots created with madmax $N_{\mathrm{plot,MM}}$ = 402,154,43.207 plots, and number of plots created with the standard plotter $N_{\mathrm{plot,std}}$ = 134,051,47.736 plots.
\[
  N_{\mathrm{plot,C5}} = \frac{S_{\mathrm{C5}}}{S_{\mathrm{plot,C5}}}
\]
\[
  N_{\mathrm{plot,MM}} = \frac{S_{\mathrm{MM}}}{S_{\mathrm{plot}}},
  \quad
  N_{\mathrm{plot,std}} = \frac{S_{\mathrm{std}}}{S_{\mathrm{plot}}}.
\]
 \\
Number of nodes that plot compressed plots $N_{\mathrm{node,C5}}$ = 150,000 nodes and uncompressed plots $N_{\mathrm{node,uncompressed}}$ = 100,000 nodes :
\[
  N_{\mathrm{node,C5}} = N_{\mathrm{node}} \, f_{\mathrm{BB,enter}}
  \quad
\]
\[
  N_{\mathrm{node,uncompressed}} = N_{\mathrm{node}} - N_{\mathrm{node,C5}}
  \quad
\]
\subsubsection{Plotting Energy}
Energy needed to create C5 compressed plots with BladeBit Ramplot $E_{\mathrm{plot,C5,RAM}}$ = 13126674.470 kWh,  $E_{\mathrm{plot,C5,GPU}}$ = 6812934.012 kWh.
\\
Energy needed to create uncompressed plots with Madmax $E_{\mathrm{plot,MM}}$ = 58942235.661 kWh and with Chia's standard plotter $E_{\mathrm{plot,std}}$ = 105794766.444 kWh.
\begin{equation}
    \begin{split}
        E_{\mathrm{plot,C5,RAM}}
    &= N_{\mathrm{plot,C5}} \, E_{\mathrm{plot,C5,RAM}} \times 0.5 \times \, \mathrm{PUE}
    \end{split}
\end{equation}
\begin{equation}
    \begin{split}
        E_{\mathrm{plot,C5,GPU}}
    &= N_{\mathrm{plot,C5}} \, E_{\mathrm{plot,C5,GPU}} \times 0.5 \times \, \mathrm{PUE}
    \end{split}
\end{equation}
\begin{equation}
    \begin{split}
        E_{\mathrm{plot,MM}} = N_{\mathrm{plot,MM}} \, E_{\mathrm{plot,MM}} \, \mathrm{PUE}
    \end{split}
\end{equation}
\begin{equation}
    \begin{split}
        E_{\mathrm{plot,std}}
    &= N_{\mathrm{plot,std}} \, E_{\mathrm{plot}} \, \mathrm{PUE}
    \end{split}
\end{equation}
\subsubsection{Farming Energy}
$E_{\mathrm{farm}}$ = 2670706785 kWh is the yearly energy needed to farm plots.
\begin{equation}
    \begin{split}
        E_{\mathrm{farm}} = N_{\mathrm{node}} \, E_{\mathrm{farm}} \, \mathrm{PUE}
    \end{split}
\end{equation}
\subsubsection{Total Operational Energy}
$E_{\mathrm{op}}$ = 2855383395.587 kWh is the total operational energy needed for plotting and farming Chia during one year.
\begin{equation}
\begin{split}
    E_{\mathrm{op}}
  = E_{\mathrm{plot,C5,RAM}} \;+\; E_{\mathrm{plot,C5,GPU}} \;+\; E_{\mathrm{plot,MM}} \\ \;+\; E_{\mathrm{plot,std}} \;+\;E_{\mathrm{farm}}.
\end{split}
\end{equation}
\subsubsection{Electricity Carbon Emissions}
$C_{\mathrm{elec}}$ = 1096467.224 t CO\textsubscript{2} is the total carbon emissions caused by Chia's yearly operational energy needs.
\begin{equation}
  C_{\mathrm{elec}}
  = I_{\mathrm{elec}} \, E_{\mathrm{op}}
  \label{eq:elec1}
\end{equation}
\subsubsection{Embodied Carbon of Storage}
$C_{\mathrm{emb,SSD}}$ = 5688.900 t CO\textsubscript{2} is the total embodied carbon emissions caused by plotting for a year.
\begin{equation}
    \begin{split}
        C_{\mathrm{emb,SSD}} = (T_{\mathrm{writes}} \, N_{\mathrm{plot,std}} + T_{\mathrm{writes,MM}} \, N_{\mathrm{plot,MM}} \\  + T_{\mathrm{writes,BB}} \, N_{\mathrm{plot,C5}})  \, \frac{\gamma_{\mathrm{SSD}}}{\mathrm{TBW}_{\mathrm{SSD}}}
    \end{split}
\end{equation}
$C_{\mathrm{emb,GPU}}$ = 30150 t CO\textsubscript{2} is the embodied carbon emissions (CAPEX) of server machines with a GPU (for nodes that use BladeBit)
\begin{equation}
    \begin{split}
        C_{\mathrm{emb,GPU}} = N_{\mathrm{node,C5}} \times ( \gamma_{\mathrm{enter}} \, \\ + \gamma_{\mathrm{GPU}}) \times \frac{f_{\mathrm{allocation}}}{L_{\mathrm{lifetime}}}
    \end{split}
\end{equation}
$C_{\mathrm{emb,noGPU}}$ = 16750 t CO\textsubscript{2} is the embodied carbon emissions (CAPEX) of server machines without a GPU (for nodes that use BladeBit in RAM mode or that create uncompressed plots, either with MadMAx or Chia's standard plotter)
\begin{equation}
    \begin{split}
        C_{\mathrm{emb,noGPU}} = N_{\mathrm{node,uncompressed}} \times\gamma_{\mathrm{enter}} \, \\  \times \frac{f_{\mathrm{allocation}}}{L_{\mathrm{lifetime}}}
    \end{split}
\end{equation}
$C_{\mathrm{emb,HDD}}$ = 177453.13792 t CO\textsubscript{2}  is the total embodied emissions (CAPEX) caused by storing plots on HDDs for a year.
\begin{equation}
    \begin{split}
        C_{\mathrm{emb,HDD}} = S_{\mathrm{net}} \, \frac{\gamma_{\mathrm{HDD}}}{L_{\mathrm{lifetime}}}
    \end{split}
\end{equation}
\subsubsection{Total embodied carbon}
$C_{\mathrm{emb}}$ = 230042.037 t CO\textsubscript{2} is the total CO\textsubscript{2} emitted during a year due to Chia's embedded carbon.
\begin{equation}
\begin{split}
    C_{\mathrm{emb}} = C_{\mathrm{emb,SSD}} + C_{\mathrm{emb,GPU}} + C_{\mathrm{emb,noGPU}} + C_{\mathrm{emb,HDD}}
\end{split}
\end{equation}
\subsubsection{Total Carbon}
\textbf{$C_{\mathrm{total}}$ = 1.32 Mt CO\textsubscript{2}} is the total carbon emissions due to the Chia blockchain during a year.
\begin{equation}
  C_{\mathrm{total}}
  = C_{\mathrm{elec}} + C_{\mathrm{emb}}.
\end{equation}

\subsection{Second method} \label{subsection:method2}
The second method extends the first approach by accounting for hardware heterogeneity across the network. Instead of treating all nodes uniformly, we stratify farmers into three distinct cohorts: servers (15\%), desktops (60\%), and laptops (25\%), based on typical farming setups \footnote{\url{https://docs.chia.net/reference-client/farming/reference-farming-hardware/}}. Each cohort holds a share of the netspace's plots (servers 65\%, desktops 30\%, and laptops 5\%, due to inequalities in storage for different hardware). Each cohort features unique characteristics: servers use datacenter-grade infrastructure (PUE=1.58), desktops represent consumer-grade hardware (PUE=1.2), and laptops utilize personal devices (PUE=1.0). Plot compression adoption varies per cohort, with BladeBit predominantly used by servers (60\%) and some desktops (20\%), MadMax mainly used by desktops (40\%) and some laptops (15\%) which rely primarily on standard plotting (85\%). Operational energy calculations incorporate cohort-specific plotting times (sourced from \footnote{\url{https://github.com/Chia-Network/chia-blockchain/discussions/6111},\url{https://xch.farm/plotting/}} -- desktop: BladeBit 0.25h, MadMax 1.5h, standard 8h; laptop: MadMax 2h, standard 10h) and farming/plotting power profiles (from \cite{eta,laptop} -- desktop: 800 W plotting, 66 W farming; laptop: 100 W plotting, 32 W farming ), while embodied carbon includes device-specific manufacturing emissions for GPUs, RAM (0.6 kg CO\textsubscript{2}), storage, and chassis (sourced from \cite{pr13020595,su17104455,RAM}). Equations maintain the same structure as Method 1 \ref{subsection:first method} but operate on partitioned netspace and incorporate cohort-specific parameters. This cohort model refines emission estimates by accommodating hardware-dependent variations in efficiency and resource utilization absent in the homogeneous first method ; the final result is \textbf{0.884 Mt CO\textsubscript{2}}. More details are available at the master's thesis or the GitHub repository \footnote{\url{https://github.com/Soraya2972002/chia_energy_pfe.git}} containing all code used.

\subsection{Sensitivity Analysis} \label{subsec:sensitivity}
Due to limited data on node types and plot compression rates, we performed a sensitivity analysis to explore the impact of these uncertainties. By testing a range of different configurations, we aim to better understand the bounds of Chia’s environmental impact under varying assumptions. We tested three additional scenarios beyond Methods 1–2, detailed in Table \ref{table:sensitivity}: The first scenario assumes a homogeneous server network without plot compression; the second one introduces hardware diversity but still no compression ; and the third scenario lowers the share of server activity in a cohort diverse network to test for energy distribution shifts.
\begin{table*}[htbp]
\centering
\caption{Sensitivity Analysis Summary}
\label{table:sensitivity}
\begin{tabular}{|l|c|l|}
\hline
\textbf{Scenario} & \textbf{Emissions} & \textbf{Key Variation} \\
& \textbf{ (Mt CO\textsubscript{2})} & \\
\hline
\hline
Homogeneous servers   & 1.401 & All nodes = servers, no BladeBit \\
(No compression) & & \\
\hline
Tiered cohorts & 0.584 & Hardware diversity without C5 plots \\
 (No compression) & & \\
 \hline
Tiered cohorts   & 0.656 & $\downarrow$ server plot share (65\% $\to$ 30\%) \\
with compression & &\\
and lower server activity & &\\
\hline
Homogeneous servers with compression  & 1.32 & All nodes = servers, with BladeBit \\
-- Method 1 & & \\
\hline
\textbf{Tiered cohorts with compression } & \textbf{0.884} & \textbf{Baseline} \\
\textbf{-- Method 2} &&\\
\hline
\end{tabular}
\end{table*}

\section{Discussions}

Our two complementary modeling approaches yield annual carbon emissions estimates for Chia’s PoST network of approximately :

\begin{equation}
\boxed{
\begin{array}{l}
C_{\mathrm{total,\,Method1}} = 1.32\ \mathrm{Mt\,CO_2/yr} \\
C_{\mathrm{total,\,Method2}} = 0.884\ \mathrm{Mt\,CO_2/yr}
\end{array}
}
\end{equation}
\noindent\textbf{Putting 0.884 Mt CO\(_2\)/yr into perspective:}
\begin{itemize}
  \item Equivalent to the annual emissions of about 192,173 average passenger cars \cite{epa}.
  \item Roughly the same as the total annual fossil‑fuel CO\(_2\) emissions of Lesotho (0.88 Mt), Somalia (0.87 Mt) or Burundi (0.84 Mt) \cite{countries}.
\end{itemize}
By contrast, Chia’s own published figure of 0.13 TWh/yr \cite{chiapower} corresponds to only \(\approx 0.0499\ \mathrm{Mt\,CO_2/yr}\)
Thus, our homogeneous scaling (Method 1) produces emissions about \textbf{27× higher} than Chia’s claim, while our cohort‐based model (Method 2) still exceeds it by roughly \textbf{18×}. These gaps stem from Chiapower’s omission of embodied emissions in storage and hardware manufacturing and replacement, oversimplified PUE assumptions, and neglect of hardware heterogeneity.

\begin{figure*}[!t]
  \centering
  \includegraphics[width=\textwidth]{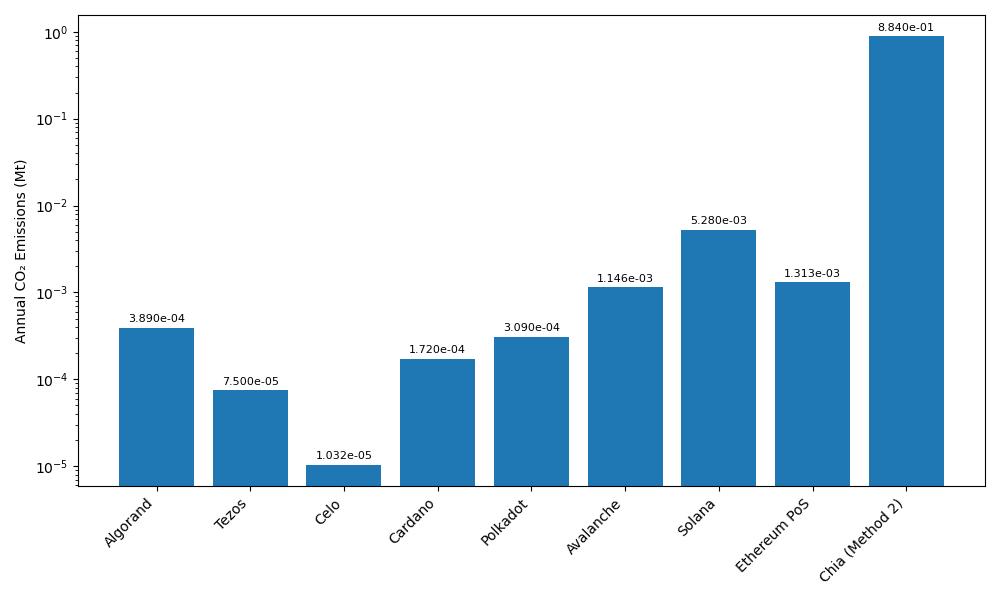}
  \caption{Comparison of annual CO$_2$ emissions across blockchains (log scale).}
  \label{fig:comparison}
\end{figure*}

While Chia’s estimated footprint of 0.884 Mt CO\textsubscript{2}/yr is orders of magnitude lower than Bitcoin’s 92.2 Mt CO\textsubscript{2}/yr, it remains significantly higher than that of other so‑called “green” blockchains\cite{ccri}. Algorand emits approximately 0.000389 Mt CO\textsubscript{2}/yr, Tezos about 0.000075 Mt CO\textsubscript{2}/yr,
Celo approximately 0.000010319 Mt CO\textsubscript{2}/yr, Cardano around 0.000172 Mt CO\textsubscript{2}/yr, Polkadot roughly 0.000309 Mt CO\textsubscript{2}/yr, Avalanche around 0.001146367 Mt CO\textsubscript{2}/yr, Solana about 0.005279701 Mt CO\textsubscript{2}/yr, and Ethereum PoS (post‑merge) about 0.001313 Mt CO\textsubscript{2}/yr. Even our lowest sensitivity scenario (0.584 Mt CO\(_2\)/yr), which reduces our estimate by roughly 35 \% relative to homogeneous scaling (Method 1), exceeds these by more than two orders of magnitude. This gap stems from PoST’s reliance on large‑scale storage—shifting emissions from computation to storage hardware production, SSD wear, and farming infrastructure. Figure~\ref{fig:comparison} summarizes these footprints.

\subsection{Sensitivity and Real-World Variability}
Our sensitivity analysis (Section~\ref{subsec:sensitivity}) reveals Chia's footprint could range from 0.584–1.402~Mt~CO\textsubscript{2}/yr. Embodied carbon accounts for 17–66\% of total emissions, underscoring that hardware manufacturing and replacement are non-negligible contributors to Chia’s carbon footprint. The differences stem from:
\begin{enumerate}
    \item \textbf{Hardware distribution} significantly impacts results: Shifting netspace toward desktops/laptops (lower PUE/embodied carbon) reduces emissions by 25\% (0.884~Mt → 0.657~Mt).
    \item \textbf{Compression trade-offs:} BladeBit requires high-resource hardware (GPUs/RAM), increasing embodied carbon by 43.14\% in the cohort-based scenarios.
    \item \textbf{Server bias inflates estimates:} Method 1 (server-only) overestimates emissions by 49.3\% compared to Method 2, highlighting the need for hardware stratification.
\end{enumerate}
\subsection{Addressing Chia's Sustainability Claims}
Chia's environmental arguments require critical reassessment:
\begin{itemize}
    \item \textbf{"Farming is negligible":} While per-node farming power is low, continuous 24/7/365 operation across 250k nodes creates substantial cumulative energy demand
    \item \textbf{"Storage would exist anyway":} Ignores accelerated replacement cycles induced by plotting wear (more than 1.6 TiB writes/plot for standard plotting) and specific storage and hardware bought by competitive farmers
    \item \textbf{"One-time plotting cost":} Fails to account for perpetual netspace growth driven by competitive farming incentives (12.7 EiB/year in our study)
\end{itemize}

\section{Conclusion and Future Work}

We have presented the first combined empirical–theoretical assessment of Chia’s environmental impact. By measuring per‐plot and per‐node energy and I/O costs on a controlled testbed, then scaling via five modeling frameworks, we show that Chia’s annual carbon footprint lies between 0.584 Mt and 1.402 Mt CO\(_2\)/y, far above the network’s own claims (0.05 Mt) and two orders of magnitude larger than mainstream PoS blockchains. Following \cite{SAI2024100169} \cite{LEI2021112422}'s guidelines, we provide complete source code at \url{https://github.com/Soraya2972002/chia_energy_pfe.git}.

This study opens several avenues for further research:
\begin{itemize}
  \item \textbf{Geographic and Grid‐mix analysis:} Break down emissions by region using localized carbon‐intensity factors, to reflect where plotters and farmers actually operate.
  \item \textbf{Broader empirical coverage:} Conduct measurements on a wider variety of hardware configurations, and collect accurate statistics on real‐world adoption.
  \item \textbf{Life‐cycle assessment (LCA):} Extend beyond storage and compute to include network‐equipment (routers, switches) and end‐of‐life recycling.
  \item \textbf{Transaction‐level impact:} Amortize network emissions per transaction or smart‐contract execution to compare operational sustainability with other blockchains on a use‐case basis.
\end{itemize}

By combining rigorous measurement, transparent modeling, and scenario analysis, future work can further guide sustainable design in blockchain protocols.

\bibliographystyle{plain}
\bibliography{references}

\end{document}